\documentstyle[11pt,epsf,paspconf]{article}

\newcommand{\lya}{Ly$\alpha$\ }
\newcommand{\nh}{N_{\rm HI}}
\newcommand{\kms}{{\rm km}\,{\rm s}^{-1}}
\newcommand{\fn}{f(N_{\rm HI})}

\newcommand\cdunits{{\rm cm}^{-2}}

\begin{document}

\title{Analysis of the Lyman-alpha Forest in Cosmological
Simulations Using Voigt-Profile Decomposition}

\author{Romeel Dav\'e and Lars Hernquist}
\affil{Astronomy Department, University of California,
    Santa Cruz, CA 95064}

\author{David H. Weinberg}
\affil{Astronomy Department, Ohio State University, Columbus, OH 43210}

\author{Neal Katz}
\affil{Astronomy Department, University of Washington, Seattle, WA 98195}

\begin{abstract}
We use an automated Voigt-profile fitting procedure to extract statistical
properties of the Ly$\alpha$ forest in a numerical simulation of an $\Omega=1$,
cold dark matter (CDM) universe.  Our analysis method is similar to that 
used in most observational studies of the forest, and we compare the
simulations to recently published results derived from Keck HIRES spectra.
With the Voigt-profile
decomposition analysis, the simulation reproduces the
large number of weak lines ($N_{\rm HI}\la 10^{14}\cdunits$) found
in the HIRES spectra.  
At $z=3$, the $b$-parameter 
distribution has a median of $35\;\kms$ and a dispersion of $20\;\kms$,
in reasonable agreement with the observed values.
The comparison between our new analysis and recent data strengthens
earlier claims that the \lya forest arises naturally in hierarchical
structure formation as photoionized gas falls
into dark matter potential wells.  

\end{abstract}

\keywords{quasars: absorption lines --- large-scale structure of universe
--- galaxies: formation}

\section{Introduction}

Absorption lines in quasar spectra,
especially the ``forest'' of \lya lines produced by concentrations
of neutral hydrogen, are uniquely suited
for probing structure formation in the high-redshift universe.
Recent cosmological simulations that incorporate gas dynamics, radiative
cooling, and photoionization reproduce many of the observed
features of quasar absorption spectra, suggesting that the \lya forest
arises as a natural consequence of hierarchical structure formation
in a universe with a photoionizing UV background
(Cen et al. 1994; Zhang, Anninos, \& Norman 1995; Hernquist et al.\ 1996, 
hereafter HKWM).
Meanwhile, high-precision observations made using the HIRES spectrograph on the
10m Keck telescope have quantified the statistics of the low column density 
absorbers to
unprecedented accuracy (e.g., Hu et al. 1995, hereafter HKCSR).
Most of the lines found in HIRES spectra are weak absorbers with
column densities $\nh < 10^{14}\cdunits$.
In order to compare to published line population statistics from
HIRES data, it is essential to analyze the simulated
spectra by Voigt-profile decomposition.
Here we compare simulated \lya spectra 
from a simulation of the cold dark matter (CDM) scenario
to HKCSR using an automated Voigt-profile fitting algorithm.
The results presented here are a condensed version of those in 
Dav\'e et al. (1996).

The physical model implicit in the decomposition technique
is that of a collection of discrete, compact clouds,
each characterized by a single velocity dispersion (thermal and/or turbulent).
The simulations undermine this physical picture because
the absorbing systems merge continuously
into a smoothly fluctuating background, often contain gas at a range
of temperatures, and are usually broadened in frequency space by
coherent velocity flows that do not resemble Gaussian turbulence.
Nonetheless, any spectrum can be described phenomenologically by
a superposition of Voigt-profile lines, with the number of components
increasing as the signal-to-noise ratio improves and more subtle
features must be matched.  The distributions of fitted column densities
and $b$-parameters provide a useful statistical basis for
comparing simulations and observations, and this is the approach that
we adopt in these proceedings.  

\section{Simulation and Artificial Spectra}

The simulation analyzed here has the same initial conditions, cosmological
parameters, and numerical parameters as that of HKWM (the reader is referred
there for more details):
a CDM universe with $\Omega=1$, 
$H_0 = 50$~km~s$^{-1}$Mpc$^{-1}$, baryon fraction $\Omega_b=0.05$, 
$\sigma_8=0.7$, (roughly the value
required to reproduce observed galaxy cluster masses; 
White, Efstathiou, \& Frenk 1993),
in a periodic simulation cube 22.222 comoving Mpc on a side.
We generate artificial spectra at $z=2$ and $z=3$ along 300 random
lines of sight through the simulation cube, using the
methods described in HKWM and Cen et al.\ (1994).

Instead of the $\nu^{-1}$ UV background spectrum adopted by HKWM,
we use the spectrum of Haardt \& Madau (1996; hereafter HM),
which is computed as a function of redshift based on the UV output
of observed quasars and reprocessing by the observed \lya forest.
The mean opacity of the \lya forest depends on the parameter combination
$\Omega_b^2/\Gamma$. 
Since observational determinations of $\Omega_b$ and $\Gamma$
remain quite uncertain, we treat the overall intensity of the UV
background as a free parameter and scale it to match the
mean \lya optical depth of Press, Rybicki, \& Schneider 
(1993; hereafter PRS) of $\bar\tau_\alpha = 0.0037(1+z)^{3.46}$.
In order to do this at $z=3$ with the original HM background intensity, we
require $\Omega_b \approx 0.08$, in better agreement with
Tytler, Fan, \& Burles (1996).
Once $\Omega_b^2/\Gamma$ is set,
there is no further freedom to adjust the simulation predictions,
and the remaining properties of the \lya forest provide tests of
the cosmological scenario.

\section{Fitting Voigt Profiles to Artificial Spectra}

We want the analysis of our simulated spectra to closely match
that used in typical observational studies, HKCSR in particular.
To this end, we have developed an automated Voigt-profile fitting routine,
AUTOVP, which allows us to efficiently handle large quantities of
simulated data and which provides an objective algorithm that
can be applied to real data.  
We first add noise ($S/N=50$) to our simulated spectra.
We then estimate the continuum in the simulated spectra,
neglecting our {\it a priori} knowledge of it.
Finally we apply AUTOVP to detect lines and fit Voigt profiles.  
In its first phase, AUTOVP identifies lines and makes an initial
estimate of their column densities and $b$-parameters,  
and in its second phase, AUTOVP takes the initial guess 
and performs a simultaneous $\chi^2$-minimization on the parameters
($v_{\rm central}, N_{\rm HI}, b$) of all lines within each detection region.  
Three independent minimization techniques are employed in conjunction
in order to reliably identify the global $\chi^2$ minimum, and lines
are added or removed based on formal significance criteria.

\section{Results}

\begin{figure}
\caption{
LHS: Column density distributions $\fn$, the number
of lines per unit redshift per linear interval of HI column density.
Error bars denote $1\sigma$ Poisson counting errors.
RHS: Distributions of $b$-parameters, for lines with 
$N_{\rm HI}>10^{13}\;\cdunits$.  } \label{fig: col}
\centerline{
\epsfxsize=3.0truein
\epsfbox[120 560 470 760]{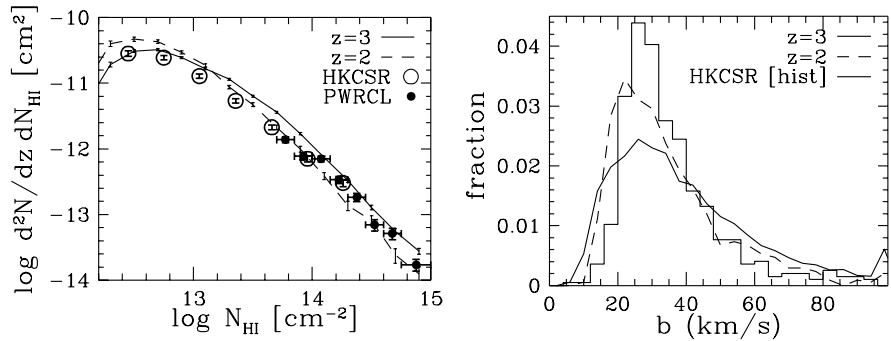}
}
\end{figure}

Figure~\ref{fig: col} shows the column density distribution 
$\fn$, the number of lines per unit
redshift per linear interval of $\nh$.
Solid and dashed lines show the simulation results from AUTOVP
at $z=3$ and $z=2$, respectively.
Filled and open circles in Figure~\ref{fig: col} show the observational
results of PWRCL and HKCSR, respectively.
We compute the HKCSR $\fn$ directly from their published line list,
with no corrections for ``incompleteness.''

The mean redshift of the HKCSR lines is $\bar z=2.9$, so the
closest comparison is to the $z=3$ simulation results.
To make this comparison more exact, we convolved the $z=3$ artificial
spectra to a resolution
$\Delta\lambda = 0.06$\AA\ ($\Delta v=3.7\;\kms$) before applying AUTOVP.
When analyzed by Voigt-profile decomposition, 
the simulation overproduces the number of lines by a factor of $1.5-2$
in the column density range
$10^{13}\;\cdunits\la N_{\rm HI}\la 10^{14}\;\cdunits$.
This excess of lines may therefore indicate a failure
of the $\Omega=1$, $\sigma_8=0.7$ CDM model.
An alternative possibility, quite plausible at present, is that
we have set the intensity of the UV background too low given our
adopted value of $\Omega_b$.  
If we adjust our intensity to force agreement with HKCSR's $\fn$ distribution,
this lowers the mean optical depth to $\bar\tau_\alpha\approx 0.32$.
This is well outside the $1\sigma$ range of PRS (figure~4),
but is somewhat {\it above} the value 
$\bar\tau_\alpha(z=3) \sim 0.25$ found by Zuo \& Lu (1993).
The uncertainty of our conclusions highlights the need for better
observational determinations of $\bar\tau_\alpha(z)$;
if $\bar\tau_\alpha$ is
well known then the amplitude of $\fn$ becomes an 
independent test of the high-redshift structure predicted by
a cosmological model.

Figure~\ref{fig: col}(b) shows the distribution of $b$-parameters
for lines with $\nh \geq 10^{13}\;\cdunits$ from HKCSR
(solid histogram) and from the AUTOVP analyses of the simulation at $z=3$
and $z=2$ (solid and dashed curves, respectively).
We only use lines with $\nh \geq 10^{13}\;\cdunits$, which eliminates lines 
whose identification
and derived properties are sensitive to the value of $S/N$ or to
details of the fitting procedure.  We find that distribution mean, median,
and dispersion are 34.6, 39.3, and 20.8 $\kms$, respectively.  From HKCSR
line lists we obtain corresponding values of 31.4, 35.8, and 16.3 $\kms$;
while systematically lower,
the agreement is reasonable given that the analysis procedures
are not identical in all their details.  The most significant difference
in the distributions is the presence of more narrow ($b<20\;\kms$)
lines in the simulation than in the data.   
A possible explanation is that our equilibrium treatment of photoionization
suppresses heating that can occur during rapid reionization (Miralda-Escud\'e
\& Rees 1994).


In summary, a Voigt-profile decomposition of simulated spectra in a
CDM universe reproduces the column density and $b$-parameter
distributions from Keck data reasonably well.
Sharper tests of cosmological models against the statistics of the \lya
forest can be obtained by expanding the redshift range of comparisons,
by improving the determination of $\bar\tau_\alpha(z)$, and by
applying AUTOVP to observational data, so that the analyses of simulated
and observed spectra are identical in detail.  
More interestingly, we are investigating different characterizations
of line profiles which more accurately describe the physical state of the gas.
While the spectra can always be fit within a 
discrete ``cloud'' model by postulating just the right
clustering properties, the ubiquitous asymmetries and non-Gaussian features
in the line profiles more likely signify the breakdown of
the Voigt-profile paradigm itself, revealing the origin of the \lya
forest in the diffuse, undulating gas distribution
of the high-redshift universe.


\begin{references}
\reference Cen, R., Miralda-Escud\'e, J.,
    Ostriker, J.P., \& Rauch M. 1994, \apj, 427, L9
\reference Dav\'e, R., Hernquist, L.H., Katz, N. \& 
    Weinberg, D.H., 1996, \apj, submitted
\reference Haardt, F. \& Madau, P. 1996, \apj, 461, 20 (HM)
\reference Hernquist, L.H., Katz, N., Weinberg, D.H., 
    \& Miralda-Escud\'e, J. 1996, \apjl, 457, L51 (HKWM)
\reference Hu, E.M., Kim, T.S., Cowie, L.L.,
    Songaila, A., \& Rauch, M. 1995 \aj, 110, 1526 (HKCSR)
\reference Miralda-Escud\'e, J. \& Rees, M. 1994, \mnras, 266, 343
\reference Petitjean, P., Webb, J.K.,
    Rauch, M., Carswell, R.F, \& Lanzetta, K. 1993, \mnras, 262, 499 
    (PWRCL)
\reference Press, W.H., Rybicki, G.B., \&
    Schneider, D.P. 1993, \apj, 414, 64 (PRS)
\reference Tytler, D., Fan, X.M.,
    \& Burles, S. 1996, Nature, 381, 207
\reference White, S. D. M.,
    Efstathiou, G., \& Frenk, C. S. 1993, \mnras, 262, 1023
\reference Zhang, Y., Anninos, P.,
    \& Norman, M.L. 1995, \apjl, 453, L57
\reference Zuo, L. \& Lu, L. 1993, \apj, 418, 601
\end{references}
\end{document}